\documentstyle[amssymb,aps,floats,twocolumn,epsf,psfig]{revtex}
\begin{document}
\draft
\title{Quartett formation at (100)/(110)-interfaces of $\mbox{\boldmath$d$}$-wave 
superconductors}
\author{A. A. Aligia$^a$, A. P. Kampf$^b$, and J. Mannhart$^b$}
\address{\sl 
$^a$Centro At\'omico Bariloche and Instituto Balseiro,\\
Comisi\'on Nacional de Energ\'ia At\'omica, 8400 Bariloche, Argentina\\
$^b$Institute of Physics, Center for Electronic Correlations and Magnetism,\\
University of Augsburg, 86135 Augsburg, Germany}
\address{~
\parbox{14cm}{\rm 
\medskip
\vskip0.2cm
Across a faceted (100)/(110) interface between two 
$d_{x^2-y^2}$-superconductors the structure of the superconducting order 
parameter leads to an alternating sign of the local Josephson coupling. 
Describing the Cooper pair motion along and across the interface by a 
one-dimensional boson lattice model, we show that a small attractive 
interaction between the bosons boosts boson binding at the interface -- a 
phenomenon, which is intimately tied to the staggered sequence of 0- and 
$\pi$-junction contacts along the interface. We connect this finding to the 
recently observed $h/4e$ oscillations in (100)/(110) SQUIDS of cuprate 
superconductors.
}}
\maketitle
\tighten

The $d_{x^2-y^2}$-symmetry of the superconducting state in high-$T_c$ cuprates 
causes a wealth of new phenomena at surfaces, grain boundaries or interfaces 
in these materials. In particular, the sign change of the order parameter 
around the Fermi surface is the origin of the most compelling experimental 
evidence for the $d$-wave nature of superconductivity in cuprates, as became 
manifest in the observation of half-flux quanta at interfaces on tricrystal 
substrates \cite{Tsuei,TsueiKirtley}. Already prior to these experiments it 
was recognized that conventional Josephson junctions (0-junctions) as well 
as $\pi$-junctions with a sign reversal of the Josephson coupling 
\cite{SigristRice} can be realized in contacts between cuprate superconductors 
depending on the mutual orientation of their crystal lattice and the attached 
four-fold symmetry of the order parameters.

In (100)/(110) interfaces or grain boundaries of $d$-wave, cuprate 
superconductors the CuO$_2$ lattices meet at 45 degrees, such that the 
$d_{x^2-y^2}$-order parameter lobes of the two superconductors point from a 
nodal towards an antinodal direction (see also Fig. 1). If the interface were 
perfectly flat, no net tunneling supercurrent would therefore flow. 
Microscopic roughness, however, allows for local supercurrents across 
interface facets \cite{Mannhart96}; the current direction at each facet is 
thereby determined by the relative phase of the clover leave lobes pointing 
towards the facet's surface. This special situation at (100)/(110) interfaces 
has led to a variety of effects like spontaneous supercurrent loops
\cite{Mannhart96}, locally time-reversal symmetry breaking phases 
\cite{Covington,Sigrist}, or anomalous field dependencies of the critical 
current density \cite{Chew}. Yet another peculiar experimental observation 
was recently reported for SQUIDs with (100)/(110) interfaces, where the flux 
periodicity of the I-V characteristics was found to be h/4e, i.e. half a flux 
quantum \cite{Schneider}; this finding is the motivation for the present 
work, in which we propose a possible mechanism for pair binding or quartett 
formation in the interface.

Networks of Josephson junctions in array geometries or even granular 
superconductors are conveniently modelled by classical XY- or extended 
quantum phase Hamiltonians \cite{Doniach}. These models in fact can be 
derived from a purely bosonic description for the Cooper pair tunneling 
processes, if fluctuations in the bulk of the superconducting order parameter 
can be neglected \cite{Grinstein}. By this means the boson kinetic energy 
translates directly into the Josephson coupling energy of the quantum phase 
Hamiltonian. The boson formulation furthermore allows for the advantage, that 
in the hard-core limit an exact mapping to a spin-1/2 Hamiltonian is possible 
\cite{Fisher}, so that preexisting knowledge for the spin model can be 
transferred to the boson problem.

In this Letter we follow the latter strategy to analyze a bosonic lattice 
model Hamiltonian with a staggered sign for the hopping amplitude representing 
an alternating sequence of superconducting 0- or $\pi$-junctions. We show that 
the special staggered structure of the kinetic energy term strongly enhances 
the tendency towards boson pair formation in the presence of a weak attractive 
interaction, as revealed by the formation of bound triplets in the groundstate 
of the equivalent spin Hamiltonian. In a closed loop Aharonov-Bohm SQUID 
geometry of the underlying boson model, oscillations with a flux periodicity 
$h/q$ are therefore expected, where $q$ is the total charge of a boson pair, 
i.e. an electronic quartett. We interpret our results as a hint for a possible 
and intriguing alternative explanation of the observed $h/4e$ oscillations in 
high-$T_c$ SQUIDS with (100)/(110) interfaces \cite{Schneider}. 

\begin{figure}[t!]
\psfig{file=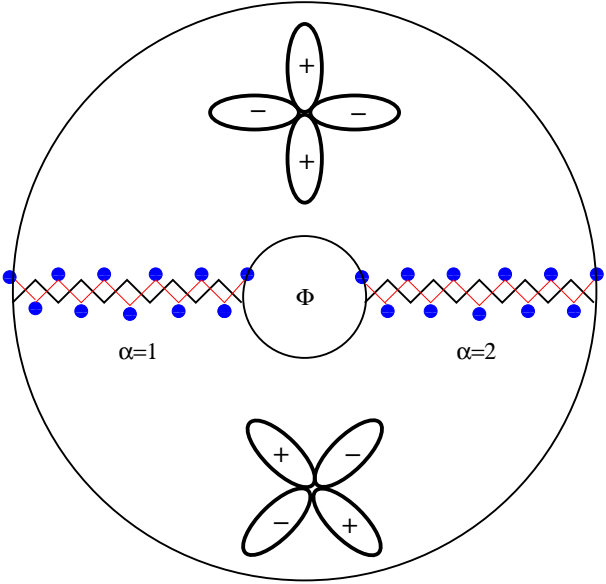,width=90mm,silent=} 
\vspace{2mm}
\caption{SQUID geometry for two $d_{x^2-y^2}$-superconductors with two 
(100)/(110) interface contact regions (labelled $\alpha=1,2$) represented as 
bold saw-tooth lines. The thin zig-zag line, which crosses the interface 
sections and connects the circles, defines the chain for the model Hamiltonian 
Eq. (1).}
\label{squidgeo}
\end{figure}

We start from the geometry shown in Fig. \ref{squidgeo} and translate it into 
the Hamiltonian  
\begin{eqnarray}
&H&=\sum_{\alpha i}\left(-t(-1)^ia^+_{\alpha i+1}a_{\alpha i}-t'a^+_{\alpha 
i+2}a_{\alpha i}+h.c.\right)+H_\Phi \nonumber \\
&+&\sum_{\alpha i}\left[Va^+_{\alpha i}a_{\alpha i}a^+_{\alpha i+1}a_{\alpha i+1}+U\left(a^+_{\alpha i}a_{\alpha i}-1\right)a^+_{\alpha i}
a_{\alpha i}\right]
\label{SQUIDHamil}
\end{eqnarray}
with boson creation and annihilation operators $a^+_{\alpha i}$ and 
$a_{\alpha i}$ and $H_\Phi=-t_\perp\sum_{j}a^+_{1j}a_{2j}e^{{\rm i}\Phi(-1)^j
/2}+h.c.$. In the disk-shape geometry in Fig. 1 the (100)/(110) interface 
between the two $d$-wave superconductors is represented by a saw-tooth line - 
assuming that the interface splits into a regular sequence of orthogonal 
facets. In a dc-SQUID setup a magnetic flux $\Phi$ may pass through the hole 
in the disk center, which separates the two interfaces labelled by 
$\alpha=1,2$. The circles mark chain sites, between which bosons (Cooper 
pairs) can hop with or without crossing the interface. The latter 
next-nearest-neighbor processes have the unique sign $-t'$ for their hopping 
amplitude, while the former processes have an amplitude with an alternating 
sign due to the misalignment by 45$^\circ$ of the $d_{x^2-y^2}$-wave order 
parameter lobes on both sides of the interface. In Eq. (\ref{SQUIDHamil}) $U$ 
and $V$ denote the onsite and nearest-neighbor interaction strengths; in the 
following we will in particular explore the effect of a weak attraction $V<0$. 
The two interfaces $\alpha=1,2$ are connected by $t_\perp$, which contains 
the phase factor of the threading flux $\Phi$. If boson (Cooper pair) binding 
occurs in the interface, oscillations with flux periodicity $h/4e$ are 
expected.

A phase change for the boson operators at every second pair of adjacent sites 
according to $b^+_{\alpha 4i}=-a^+_{\alpha 4i}$, $b^+_{\alpha 4i+1}=-a^+_{
\alpha 4i+1}$, $b^+_{\alpha 4i+2}=a^+_{\alpha 4i+2}$, $b^+_{\alpha 4i+3}=a^+_{
\alpha 4i+3}$ transforms the kinetic energy part of the Hamiltonian for each 
interface into 
\begin{equation}
H_{kin}=\sum_{\alpha i}\left[ -tb_{\alpha i+1}^{+}b_{\alpha i}+t^{\prime}b_{
\alpha i+2}^{+}b_{\alpha i}+h.c.\right] \,;  \label{Hkin}
\end{equation}
all other terms remain unchanged. Importantly, for a sequence of ordinary
0-junctions the second term in Eq. (\ref{Hkin}) appears with a negative sign.

We now focus on the physics in {\it one} interface. As anticipated above we
consider the hard-core limit $U\rightarrow \infty $, in which the boson
problem maps onto a spin-1/2 model by means of the transformation 
\cite{Fisher} 
\begin{equation}
S_{i}^{+}=(-1)^{i}b_{i}\text{, }S_{i}^{-}=(-1)^{i}b_{i}^{+}\text{, }
S_{i}^{z}={\frac{1}{2}}-b_{i}^{+}b_{i}\,.  
\label{bosonspin}
\end{equation}
The resulting spin Hamiltonian reads 
\begin{eqnarray}
H_{S}=\sum_{i}\big[ && J_{1}\left(
S_{i}^{x}S_{i+1}^{x}+S_{i}^{y}S_{i+1}^{y}+\Delta S_{i}^{z}S_{i+1}^{z}\right)
\nonumber\\
&+&J_{2}\left( S_{i}^{x}S_{i+2}^{x}+S_{i}^{y}S_{i+2}^{y}\right) \big] 
\label{spinHamil}
\end{eqnarray}
with the spin exchange coupling constants $J_{1}=2t$, $J_{2}=2t^{\prime}$,
and the anisotropy parameter $\Delta =V/2t<0$. This model -- including an
additional anisotropy of the next-nearest-neighbor exchange $J_{2}$ -- has been
studied before in the context of metamagnetic transitions \cite{Aligia}. In
particular, analytical results for pairing of magnons (bosons in the original 
language) were derived, and also a tendency to form clusters of more than two 
particles was obtained for certain parameter regimes. Because of its relevance 
for the quartett formation, we start to discuss the binding problem in the 
original bosonic language. 

From the insight into the physics of the spin-chain model we infer, that 
for $t^{\prime }>0$ the partial frustration of the kinetic energy favors the 
binding of bosons, which represent the Cooper pairs. For each total 
momentum $K$ of a pair of bosons, the bound state can be written as
\begin{equation}
|\psi_{K}\rangle =\sum_{j>0}A_j\sum_{n}e^{-{\rm i}K(n+j/2)}\, b_{n+j}^{+}
b_{n}^{+}.
\label{psi}
\end{equation}
The Schr\"{o}dinger equation for the bound state, $H_1|\psi_K\rangle=\lambda_K
|\psi_K\rangle$, with the Hamiltonian $H_1$ of one interface can be solved 
with the following ansatz 
\begin{equation}
A_j=(\gamma_{1})^{j}-(\gamma _{2})^{j},  
\label{aj}
\end{equation}
where $\gamma_1$ and $\gamma_2$ are the two solutions of the equation
\begin{equation}
-2t\cos \frac{K}{2}\left( \frac{1}{\gamma }+\gamma \right)
+2t^{\prime }\cos K\left( \frac{1}{\gamma^{2}}+\gamma^{2}\right)=\lambda_K\, ,
\label{gam}
\end{equation}
with $|\gamma_{1,2}|<1$; the eigenvalue $\lambda_K$ has to satisfy
\begin{eqnarray}
\lambda_K=V&+&2t^{\prime}\cos K(\gamma_1^2+\gamma_2^2+\gamma_1\gamma_2+1)
\nonumber\\
&-&2t\cos \frac{K}{2}(\gamma_{1}+\gamma_{2})\, .  
\label{lam}
\end{eqnarray}
The size of the pair is determined by the quantity $\xi=-1/\ln[\max(|
\gamma_{1}|,|\gamma_{2}|)]$ and decreases with increasing $V$. 

The critical interaction $V_b$ for binding is determined by the condition, that
the minimum of $\lambda_K$ with respect to all possible pair momenta $K$ equals
twice the minimum of the one-particle energy $E_k=-2t\cos k+2t^{\prime}\cos 
(2k)$. The wave vector, which leads to the minimum $E_{k}$ is $k_{\min}=0$ 
for $\alpha=t^{\prime}/t\leqslant 1/4$ and $k_{\min}=\arccos [1/(4\alpha)]$ 
for $\alpha\geqslant 1/4$. In our analysis we find that the optimum 
two-particle wave vector is $K_{\min}=0$ for $\alpha\leqslant 1/(2\sqrt{2})$ 
in agreement with previous results for the spin-chain model \cite{Aligia}, 
and $K_{\min}=2k_{\min}$ for $\alpha\geqslant 1/(2\sqrt{2})$. The results for 
the minimum attraction necessary for binding can be summarized as follows with 
$\Delta_b=V_b/2t$:
\begin{equation}
\Delta_b=\left\{\begin{array}{l@{\quad}}-\displaystyle{1+\sqrt{1
-4\alpha}\over 2}\hskip0.9cm\text{ for }\alpha=t^{\prime}/t\leqslant 1/4, \\ 
-2\alpha\hskip1.7cm \text{ for }1/4\leqslant\alpha\leqslant 1/(2\sqrt{2}), \\ 
-\displaystyle{1\over 8\alpha}(\sqrt{16\alpha^{2}-1}+1)\hskip0.1cm\text{ for }
\alpha \geqslant 1/(2\sqrt{2})\, .
\end{array}\right.
\label{delb}
\end{equation}
This function is represented by the full line in Fig. 2. Clearly, a small to
moderate attraction is enough to lead to pair binding for positive 
$t^{\prime}$, which represents the alternating sequence of 0- and 
$\pi$-junctions, particularly for small hopping amplitudes $t$ across the 
interface. Specifically, for the physically reasonable regime $t^{\prime}/t>1$ 
an attractive interaction of order $t$ is sufficient for boson-pair formation; 
the energy scale for $t$ should be determined by the Josephson coupling 
energy. Although $\xi$ is very sensitive to $V$ and diverges for $V\rightarrow 
V_b$, typical pair sizes  for $V\sim t$ and $t^{\prime}>2t$ are an order of 
magnitude larger than the size of an individual facet.

\begin{figure}[t!]
\psfig{file=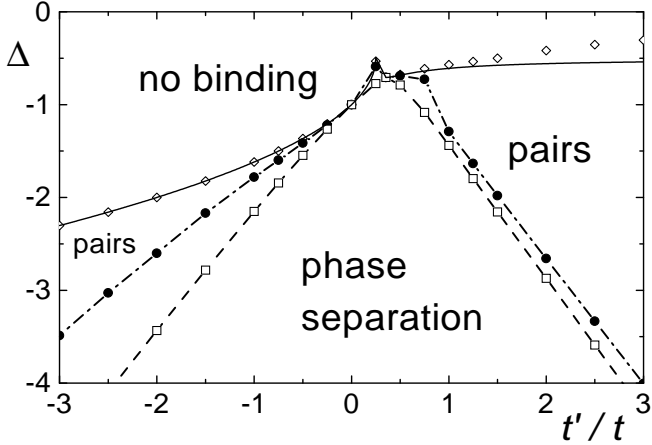,width=90mm,silent=} 
\vspace{2mm}
\caption{Phase diagram of the interface model according to the number of 
particles $n$, which form bound composites; $\Delta=V/2t$. The full line 
corresponds to the analytical solution Eq. (\ref{delb}). Open diamonds 
indicate the pair binding boundary, full circles correspond to the transition 
from $n=2$ to $n>2$, and open squares denote the onset of phase separation.} 
\label{diag}
\end{figure}

It is known, particularly in models with strong correlations, that pairing
competes with phase separation \cite{Dagotto} and the tendency to bind in 
groups of more than two particles. To explore these possibilities we have 
studied numerically the equivalent spin Hamiltonian Eq. (\ref{spinHamil}) 
in a chain of $L=16$ sites. For each total spin projection $S_z$, which 
translates into a number of flipped spins (i.e. magnons) $m=L/2-S_z$ added to 
the fully polarized ferromagnetic ground state, we have calculated the 
ground-state energy $E(m)$. To minimize finite-size effects and to accurately 
obtain the energy for one magnon, it is necessary to minimize over twisted 
boundary conditions \cite{Aligia}. If the particles in the system (bosons in 
the original language or magnons in the spin language) prefer to bind in 
groups of $n$ particles, the quantity $e(m)=(E(m)-E(0))/m$ is minimized for 
$m=n$. We argue that phase separation occurs, when the condition $E(m)>(mE(L)
+(L-m)E(0))/L$ holds for all $m$. 

In Fig. 2 we show the resulting ground-state phase diagram. $\Delta=V/2t$ is 
the measure for the strength of the attractive interaction and $\alpha=t'/t$ 
is the ratio of hopping amplitudes for the motion along and across the 
interface. $\alpha>0$ represents the alternating sequence of 0- and 
$\pi$-junctions, while $\pi$-junctions are absent for $\alpha<0$. Four 
different regions are indicated in Fig. 2: the strong attraction regime, in 
which there is phase separation, a regime without binding and two intermediate 
phases, in which the size of the optimum particle cluster is $n=2$, or $n>2$. 
In the latter region $n$ increases in unit steps as the attraction increases, 
except for $t^{\prime}>t$, where only even $n$ appear. The asymmetry between 
positive and negative $t^{\prime }$ with respect to the stability of 
boson-pair binding is evident, underlining the importance of the existence of 
$\pi$-junctions in the quartett formation. The numerical results for the 
border between $n=1$ and $n=2$ are in excellent agreement with the analytical 
results of Eqs. (\ref{delb}) -- except for $t^{\prime}>t$, where finite-size 
effects are present.

Comments remain in order about a possible origin of an attractive interaction 
for the bosons (i.e. Cooper pairs) in cuprate superconductors (or 
superconductors in general). We first note that the idea of quartett formation 
has been put forward before in nuclear physics \cite{Nozieres}; proposals 
exist, that four-particle condensation may occur as a phenomenon alternative 
or complementary to nucleon pairing \cite{Nozieres}. In cuprates, the possible 
existence of clusters of pairs has been discussed within the mesoscopic 
Jahn-Teller pairing model \cite{Muller}. It was furthermore proposed, that 
due to strong phase fluctuations cuprates may be close to an exotic 
superconducting phase with quartet condensation \cite{Hlubina}. 

A viable mechanism for electron pairing in high-T$_c$ superconductors arises 
from antiferromagnetic (AF) spin fluctuations in a doped Mott-insulating host 
\cite{Scalapino}. While a nearest-neighbor electron-electron attraction is 
dynamically generated from a local repulsive Coulomb interaction, a 
correlation of the pair motion in an environment with short range AF order is 
expected to optimally minimize the pair motion induced breaking of AF bonds. 
A simple picture for the source of binding in a system with short range AF
correlations is obtained thinking in terms of static holes added to a N\'eel
antiferromagnet on a square lattice: if two separated holes are added, they
break 8 bonds. If instead they are added as nearest neighbors, only 7 bonds
are broken. Naturally, this argument can be extended to more particles, 
suggesting that the binding mechanism may be active also for more than two 
particles and that the actual size of the composite object is determined by 
the competition with the kinetic energy of the particles, in a similar way 
as it happens in our bosonic model for a single interface.

Contrary to the few-particle problem in nuclear physics, four-particle 
interaction vertices have so far been unexplored in correlated electron 
lattice models with superconducting instabilities, and therefore no firm 
basis exists for a discussion of correlated pair motion or even quartett 
formation tendencies. Yet, it is well established, that the quantitative 
description of the spin dynamics in undoped cuprates requires to include a 
sizable ring-exchange coupling \cite{KK}, which naturally arises in 
strong-coupling expansions in next to leading order around the atomic limit 
\cite{Erwin}. A ring-exchange coupling does indeed contain 4-particle 
interactions between electrons on plaquettes of a square lattice. The 
complex structure of such an interaction has not been analyzed and its 
consequences for the dynamics of doped holes or the pair-wavefunction in 
the superconducting state remains unknown.   

In the present analysis we have assumed a small attractive pair interaction. 
It is likely that such a weak interaction does not lead to observable 
phenomena in the bulk of a correlated superconductor; but at the peculiar 
(100)/(110) interface between $d$-wave superconductors seemingly 
subdominant 4-particle correlations may lead to new pairing tendencies and 
the possibility for quartett formation. In fact, one may argue, that the 
binding tendency is enhanced at the interface, because the cost in kinetic 
energy is reduced due to the phase factors of the $d$-wave pairs and the 
concomitant staggered sign of the hopping amplitudes in the bosonic language. 

In the context of frustrated Josephson junction networks an alternative 
mechanism of Cooper-pair binding, based on a $Z_2$ symmetry of a particular 
geometry was reported for Aharonov-Bohm cages \cite{Doucot}. In this case 0- 
and $\pi$-junctions are realized on plaquettes, which are threaded by one 
flux quantum. In a one-dimensional arrangement these plaquettes are 
interconnected in a geometry, which leads to perfectly flat bands and thus to 
particle localization. Interactions may then lead to delocalized two-particle 
bound states or mobile charge $4e$ composite objects, which in closed loop 
SQUIDs should also give rise to an elementary $h/4e$ period of flux. A common 
feature of this proposal and the mechanism discussed in this Letter is indeed 
the important role of the partial frustration of kinetic energy. 

If the mechanism discussed in this letter is indeed at work at (100)/(110) 
interfaces, the experimental observation of $h/4e$ flux periodicities in 
(100)/(110) SQUIDS of high-$T_c$ superconductors would follow as a natural 
consequence. Yet, as discussed in \cite{Schneider}, more conventional 
proposals of a suppressed $\sin \varphi$ component and a dominant $\sin 2
\varphi$ component of the Josephson current at the (100)/(110) interfaces are 
available and present a vivid alternative to explain the experimental 
findings \cite{Lueck,Mints}. We believe, however, that the above discussed 
mechanism of quartett formation offers an intriguing new route for so far 
unexplored pair-binding phenomena in superconductors \cite{Ashcroft}. 

We acknowledge support by the DFG (SFB 484),  
the BMBF (EKM-project 13N6918), the ESF via the PiShift-Programme, and 
ANPCyT through PICT 03-12742. A.A.A. is partially supported by CONICET. We 
are grateful for discussions with T. Kopp, C. W. Schneider and K. Hallberg, 
and to N. Ashcroft for a hint to the work of Ref. \cite{Ashcroft}.


\begin{references}
\bibitem{Tsuei} C. C. Tsuei and J. R. Kirtley, Rev. Mod. Phys. {\bf 72}, 969 
(2000).

\bibitem{TsueiKirtley} C. C. Tsuei et al., Phys. Rev. Lett. {\bf 73}, 593 
(1995); J. R. Kirtley et al., Nature {\bf 373}, 225 (1995).

\bibitem{SigristRice} M. Sigrist and T. M. Rice, J. Phys. Soc. Jpn. {\bf 61}, 
4283 (1992).

\bibitem{Mannhart96} J. Mannhart et al., Phys. Rev. Lett. {\bf 77}, 2782 
(1996).

\bibitem{Covington} M. Covington et al., Phys. Rev. Lett. {\bf 79}, 281 (1997).

\bibitem{Sigrist} C. Honerkamp and M. Sigrist, Physica C {\bf 317}-{\bf 318}, 
489 (1999); C. Honerkamp, K. Wakabayashi, and M. Sigrist, Europhys. Lett. 
{\bf 50}, 368 (2000).

\bibitem{Chew} J. Mannhart, B. Mayer, and H. Hilgenkamp, Z. Phys. B {\bf 101},
175 (1996); N. G. Chew et al., Appl. Phys. Lett. {\bf 60}, 1516 (1997).

\bibitem{Doniach} S. Doniach, Phys. Rev. B {\bf 24}, 5063 (1981); M. P. A. 
Fisher et al., Phys. Rev. B {\bf 40}, 546 (1989).

\bibitem{Grinstein} M. P. A. Fisher and G. Grinstein, Phys. Rev. Lett. {\bf 
60}, 208 (1988).

\bibitem{Fisher} M. E. Fisher, Rep. Prog. Phys. {\bf 30}, 615 (1967).

\bibitem{Schneider} C. W. Schneider et al., Europhys. Lett. {\bf 68}, 86 
(2004). 

\bibitem{Aligia} A. A. Aligia, Phys. Rev. B {\bf 63}, 014402 (2000), 
and references therein.

\bibitem{Dagotto} See e.g. E. Dagotto and J. Riera, Phys. Rev. Lett. {\bf 
70}, 682 (1993).

\bibitem{Nozieres} G. R\"opke et al., Phys. Rev. Lett. {\bf 80}, 3177 (1998).

\bibitem{Muller} D. Mihailovic, V. V. Kabanov, and K. A. M\"uller, Europhys. 
Lett. {\bf 57}, 254 (2002).

\bibitem{Hlubina} R. Hlubina, M. Grajcar, and J. Mr\'az, preprint 
cond-mat/0304213 (unpublished).

\bibitem{Scalapino} D. J. Scalapino, Phys. Rep. {\bf 250}, 329 (1995).

\bibitem{KK} A. A. Katanin and A. P. Kampf, Phys. Rev. B {\bf 66}, 100403(R) 
(2002).

\bibitem{Erwin} E. M\"uller-Hartmann and A. Reischl, Eur. Phys. J. B {\bf 
28}, 173 (2002).

\bibitem{Lueck} E. Il'ichev et al., Phys. Rev. B {\bf 60}, 3096 (1999); Y. 
Tanaka and S. Kasiwaya, ibid. {\bf 56}, 892 (1997); T. L\"uck et al., ibid. 
{\bf 68}, 174524 (2003); T. Lindstr\"om et al., Phys. Rev. Lett. {\bf 90}, 
117002 (2003).

\bibitem{Mints} R. G. Mints, Phys. Rev. B {\bf 57}, 3221 (1998); R. Mints and 
I. Papiashvili, ibid. {\bf 64}, 134501 (2001).

\bibitem{Doucot} B. Dou\c{c}ot and J. Vidal, Phys. Rev. Lett. {\bf 88}, 
227005 (2002).

\bibitem{Ashcroft} We note, that S. Koh has extended the Gorkov decoupling 
scheme in the pairing theory of superconductivity by including pair-pair 
correlations in momentum space. See e.g. S. Koh, Physica C {\bf 191}, 167 
(1992); Phys. Rev. B {\bf 49}, 8983 (1994).
\end{references}
\end{document}